\documentclass[twoside,fleqn,a4paper]{article}

\usepackage{espcrc2MOD}
\usepackage{epsfig}
\usepackage{graphicx}
\usepackage{latexsym}
\newcommand{\AmS}{{\protect\the\textfont2
  A\kern-.1667em\lower.5ex\hbox{M}\kern-.125emS}}

\begin{document}
\sloppy
\begin{titlepage}
\title{{$F_{2}^{D(4)}$ in the Model of the Stochastic Vacuum}\thanks{Talk presented at the QCD'99 Euroconference, Montpellier, July 1999.}}
\author{Oscar E. Ram\'{\i}rez del Prado \thanks{E-mail: ramirez@thphys.uni-heidelberg.de}$^{,}$ 
\thanks{supported by the Deutscher Akademischer Austauschdienst}$^{,}$ 
\address{Institut f\"ur Theoretische Physik der Universit\"at Heidelberg,\\ Philosophenweg 16, D-69120 Heidelberg}}
\begin{abstract}
  In this paper a calculation for the diffractive structure functions
  $F_{L}^{D(4)}$ and $F_{T}^{D(4)}$ using an eikonal approximation is
  presented. A modified version of the model of the stochastic vacuum
  that includes an energy dependence is used to calculate the
  scattering amplitude of two Wilson loops which in our approach
  describe the color dipole interaction. Numerical results are
  presented in the $\beta$ region where the non-perturbative effects
  are expected to be large. The DIS diffractive cross section is also
  discussed. Our calculation for the diffractive slope in $t$ is in
  good agreement with the experiment.
\end{abstract}
\end{titlepage}
\maketitle
\section{Introduction}
Following~\cite{Ramirez1998}, we present a new approach to the
calculation of the diffractive structure functions using a modified
version of the model of the stochastic vacuum (MSV) proposed by
Rueter~\cite{Rueter1999}. He introduces an energy dependence in a
phenomenological way by assuming the existence of two pomerons which
couple to the dipoles according to their size: a hard pomeron of
exponential type which couples to relative small dipoles such that
their radius is bigger than $r_{\rm cut}$ and a soft pomeron having
the Donnachie-Landshoff type~\cite{Donnachie1984}, that couples to
dipoles of hadronic size. The radial cutoff $r_{\rm cut}= 0.16$ fm is
introduced in order to avoid tiny dipoles for which perturbative gluon
exchange has to be taken into account and the non-perturbative model
of the stochastic vacuum has to be ``switched off''.

\section{Diffractive Structure Functions}

The process under consideration is the scat\-tering of a virtual
photon off a proton: $\gamma^{*}+p \rightarrow p + {\rm X}$.  The
proton scatters elastically remaining unchanged, whereas the photon
dissociates in a color-singlet quark-antiquark pair.  We will work in
the center-of-mass frame, the natural system for the MSV.  We define
the following invariants

\begin{equation} \label{diffinvariants}
Q^2 = - q^2, \hspace{2mm} W^2=(p+q)^2, \hspace{2mm} t = (p'-p)^2
\end{equation}

\noindent where $t=-\Delta_{\perp}^2$ is the momentum transfer which is purely transversal and small: $|t|\leq 1$ \mbox{GeV$^2$}.

The kinematics of diffractive processes is usu\-ally given in terms of the variables
%\begin{equation} \label{xpomeronbeta}
$\beta = Q^2/Q^2+M^2_x$,
%\hspace{6mm} 
$x_{\mbox{\tiny I$\!$P}}=Q^2+M^2_x/Q^2+W^2$;
%\end{equation}
%\noindent 
$M_x$ is the invariant mass of the $q\bar{q}$ pair.  In a pomeron
exchange picture, $x_{\mbox{\tiny I$\!$P}}$ can be interpreted as the
fraction of the proton momentum carried by the pomeron and $\beta$ as
the momentum fraction of the struck quark within the pomeron.  The
momentum transfer and the relative momentum between the quark and the
antiquark are given respectively by
%\begin{equation} \label{delta-p}
$\vec{\Delta}_{\perp}=\vec{l}+\vec{h}$,
%\hspace{6mm} 
$\vec{p}=(1-z)\vec{l}-z\vec{h}$.
%\end{equation}
where $\vec{l}$ ($\vec{h}$) is the transverse momentum of the (anti-)quark.

The phase space written in terms of the above variables is given by

\begin{equation} \label{phasespace}
d\Phi_3=\frac{1}{(2\pi)^9}\frac{1}{16W^2}\frac{Q^2}{\beta}dz\;d|t|\;d\theta\;\frac{dx_{\mbox{\tiny I$\!$P}}}{x_{\mbox{\tiny I$\!$P}}}\;d\varphi \;,
\end{equation}

\noindent here $\theta$ is the orientation of the relative momentum between the quark and the antiquark and $\varphi$ is the orientation of the momentum transfer. 

The scattering amplitude and the cross section can be calculated
following~\cite{Ramirez1998}.  using massless quarks we get the
following expression for the longitudinal and the transversal
structure functions respectively

\begin{eqnarray} \label{fld4msv}
\lefteqn{F_{L}^{D(4)}(x_{\mbox{\tiny I$\!$P}},\beta,Q^2,t)=\frac{2}{3}\frac{8N_c}{(2\pi)^3}\frac{Q^6}{x_{\mbox{\tiny I$\!$P}}\;\beta}} \nonumber \\
&&\times \int_{z_{\rm cut}}^{1-z_{\rm cut}} dz\, z^3 (1-z)^3 \int_0^{2\pi} d\theta \nonumber \\
&& \times \left| \int_0^\infty db\, b \; {\rm J}_0(\sqrt{|t|}b) \int \frac{d^2 r}{4\pi}\; e^{-i\vec{p}\cdot\vec{r}} \right. \nonumber \\
&& \left. \hspace{1cm} {\rm K}_0(\epsilon r)\tilde{W}\left(b,\vec{r},z,W,t\right)\right|^2,
\end{eqnarray}

\begin{eqnarray} \label{ftd4msv}
\lefteqn{F_{T}^{D(4)}(x_{\mbox{\tiny I$\!$P}},\beta,Q^2,t)=\frac{2}{3}\frac{2N_c}{(2\pi)^3}\frac{Q^6}{x_{\mbox{\tiny I$\!$P}}\;\beta}} \nonumber \\
&& \times \int_{z_{\rm cut}}^{1-z_{\rm cut}} dz\, z^2 (1-z)^2 \; \left(z^2+ (1-z)^2\right)  \nonumber \\ 
&& \times \int_0^{2\pi} d\theta \; \left| \int_0^\infty db\, b \;{\rm J}_0(\sqrt{|t|}b) \int \frac{d^2 r}{4\pi}\; e^{i\theta_{r}} \right. \nonumber \\
&& \left. e^{-i\vec{p}\cdot\vec{r}}\;{\rm K}_1(\epsilon r) \tilde{W}\left(b,\vec{r},z,W,t\right)\right|^2.
\end{eqnarray}

Here, we have summed over helicities and three quark flavors, $N_c$ is
the number of quarks, ${\rm J}_0$ is the ordinary Bessel function and
${\rm K}_0$ and ${\rm K}_1$ are the 2nd. kind modified Bessel
functions of order $0$ and $1$. Here $\epsilon$ and the relative transverse
momentum of the $q\bar{q}$ pair $\vec{p}$ are given by

\begin{equation} 
\epsilon = \sqrt{z(1-z)Q^2}, \hspace{2mm} 
\vec{p} =  \epsilon \sqrt{\frac{1-\beta}{\beta}}\left(\begin{array}{c}\cos\theta\\\sin \theta\end{array}\right).
\end{equation} 

Notice that there is a cut-off $z_{\rm cut}= 0.2\;\mbox{GeV}/W$ in the
$z$-integration introduced to avoid the configurations that correspond
to $z$ near $0$ or $1$ in which the quark or the antiquark
respectively becomes slow in the center-of-mass frame and might
spoil the validity of the eikonal approximation as discussed
in~\cite{Rueter1999,DoGoPi1998}. 

\begin{eqnarray} \label{functionw}
\lefteqn{\tilde{W}(\vec{b},\vec{r},z,W,t) = \int  \frac{d^2 R}{4\pi} \left|
\Psi_P(R)\right|^2} \nonumber \\ 
&& \hspace{10mm} \times \; \tilde{J}\left(\vec{r},\vec{R},\vec{b},z\right)\,G\left(r,R,t,W\right)
\end{eqnarray}

\noindent and can be understood as the dipole-proton scattering amplitude. The proton wave function in the quark-diquark representation is given by a Gaussian function

\begin{equation} \label{protonwf}
\Psi_P(R)=\frac{\sqrt{2}}{S_P}e^{-\frac{R^2}{4S_P^2}},
\end{equation}

The function $\tilde{J}\left(\vec{r},\vec{R},\vec{b},z,W,t\right)$ is the dipole-dipole scattering amplitude evaluated in the MSV~\cite{Ramirez1998,Rueter1999}.

By performing the integration over $t$, a more inclusive and experimentally more accessible quantity can be defined

\begin{equation} \label{f2d4integrated}
F^{\rm D(3)}_{\rm 2}(x_{\mbox{\tiny I$\!$P}},\beta,Q^2) = \int{dt} \; F^{\rm D(4)}_{\rm 2}(x_{\mbox{\tiny I$\!$P}},\beta,Q^2,t)\;.
\end{equation}

Notice that in both longitudinal and transversal structure functions there is an explicit factor of the form $1/x_{\mbox{\tiny I$\!$P}}$ which is a direct consequence of the kinematics of the process, as can be seen from the expression for the phase space Eq.~(\ref{phasespace}). At first glance, one would say that our approach admits a factorization of the form

\begin{equation} \label{factorization}
F^{\rm D(3)}_{\rm 2}(x_{\mbox{\tiny I$\!$P}},\beta,Q^2) = f_p(x_{\mbox{\tiny I$\!$P}}) \; f(\beta,Q^2) 
\end{equation}

\noindent where $f_p$ is a pomeron flux factor depending only on $x_{\mbox{\tiny I$\!$P}}$. 
However, this is not the case, since the function
$\tilde{W}\left(b,\vec{r},z,W,t\right)$ depends in a non-trivial way
on $W$ due to a cut in the z range $z_{\rm cut}$ (see above).
Hence for fixed $Q^2$, the function $\tilde{W}$ depends on
$x_{\mbox{\tiny I$\!$P}}$.

\begin{center}
\begin{figure*}[!htp]
\begin{center}
\unitlength1cm
\begin{minipage}[t]{5.4cm}
\begin{center}
\begin{picture}(4.5,4.5)
\put(4.5,0){\scriptsize $x_{\mbox{\tiny I$\!$P}}$}
\put(0,0){\centerline{\epsfysize=4.5cm\epsffile{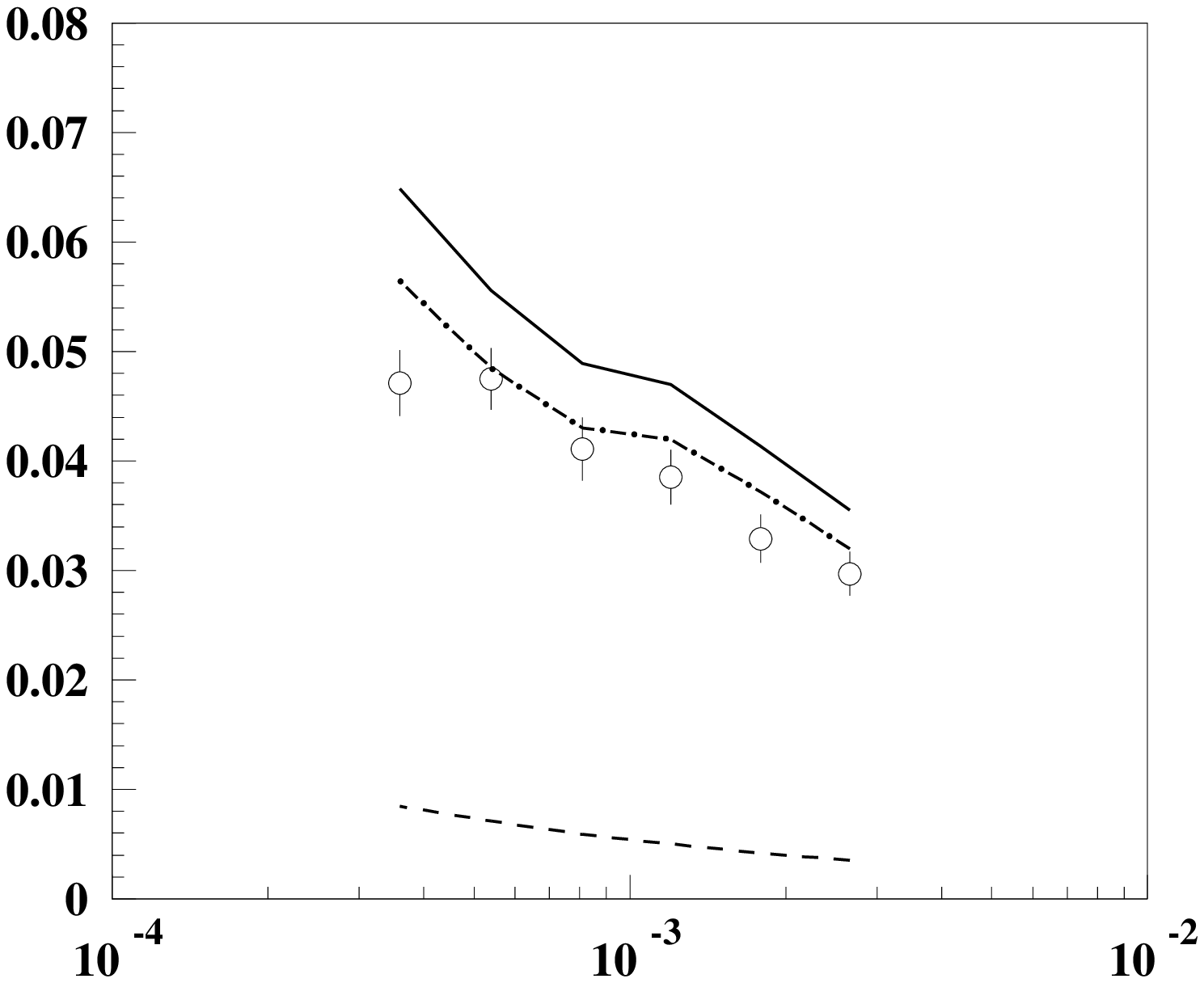}}}
\end{picture}
{\scriptsize \mbox{} \\(a) $Q^2=8.0$~\mbox{GeV$^2$}, $\beta = 0.667$}
\end{center}
\end{minipage}
\ \ \hspace{0.5cm}
\begin{minipage}[t]{5.4cm}
\begin{center}
\begin{picture}(4.5,4.5)
\put(4.5,0){\scriptsize $x_{\mbox{\tiny I$\!$P}}$}
\put(0,0){\centerline{\epsfysize=4.5cm\epsffile{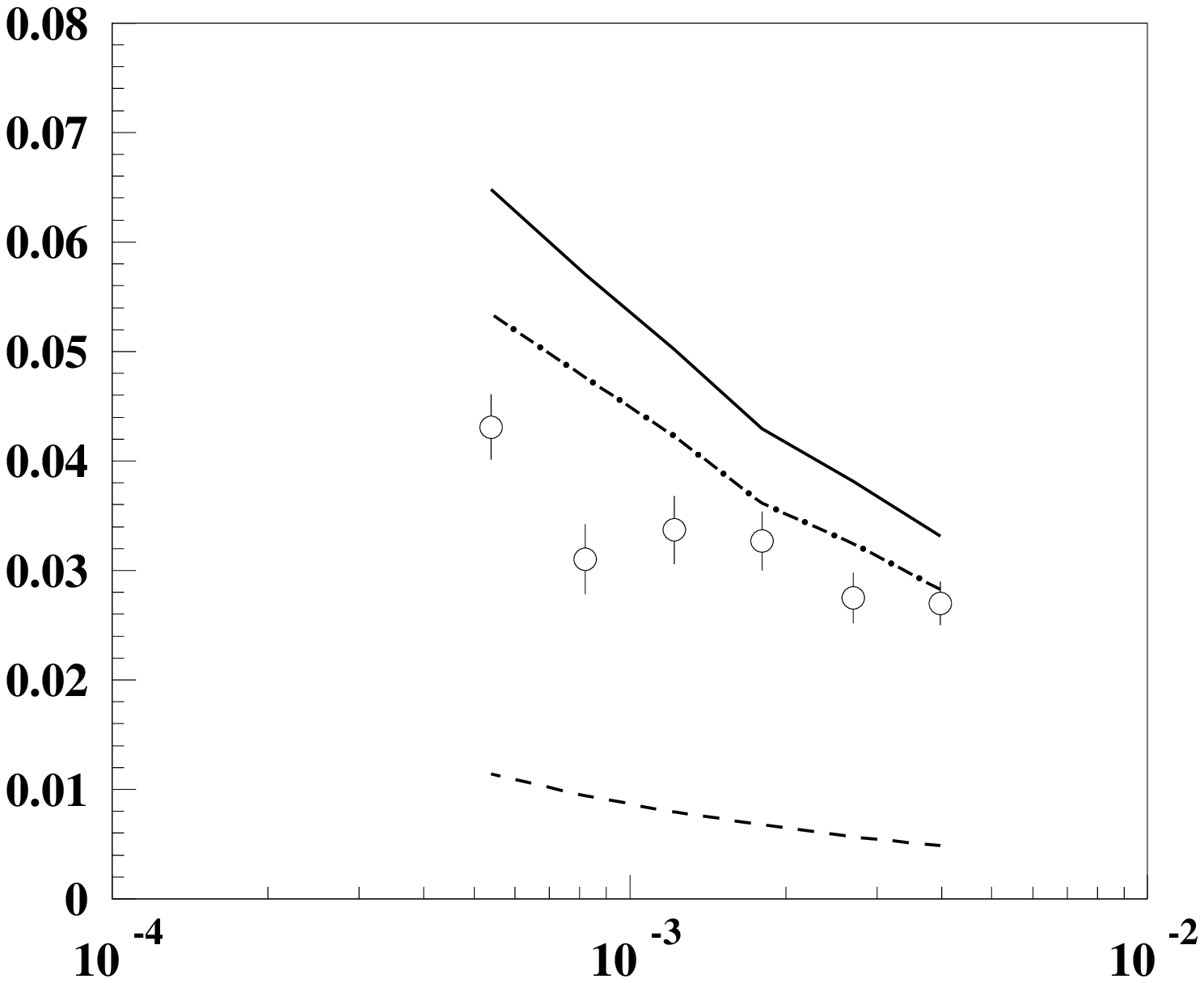}}}
\end{picture}
{\scriptsize \mbox{} \\ (b) $Q^2=14.0$~\mbox{GeV$^2$}, $\beta = 0.778$}
\end{center}
\end{minipage}
\\
\begin{minipage}[t]{5.4cm}
\begin{center}
\begin{picture}(4.5,4.5)
%\vspace{-0.8cm}
\put(4.5,0){\scriptsize $x_{\mbox{\tiny I$\!$P}}$}
\put(0,0){\centerline{\epsfysize=4.5cm\epsffile{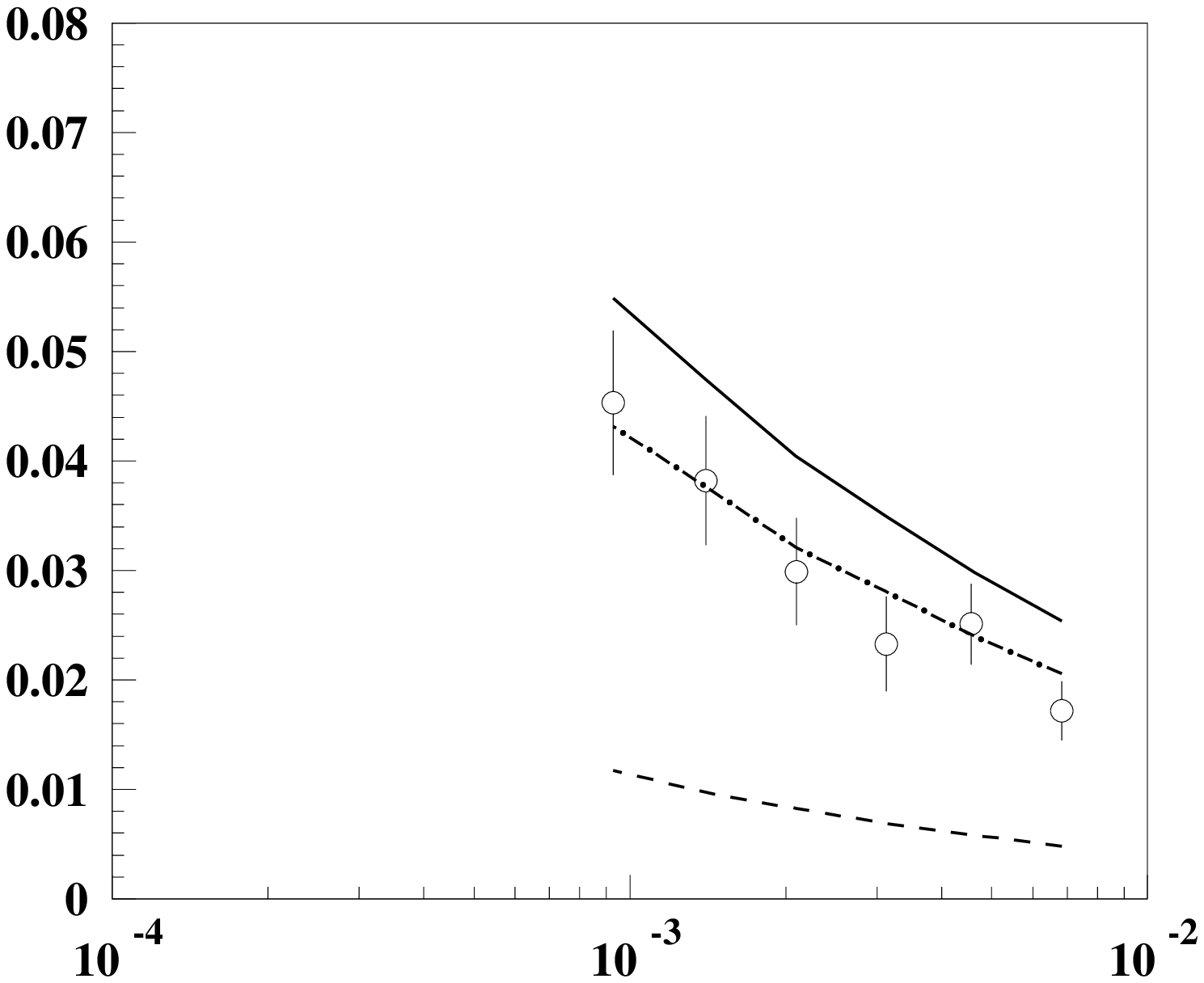}}}
\end{picture}
{\scriptsize \mbox{} \\ (c) $Q^2=27.0$~\mbox{GeV$^2$}, $\beta = 0.871$}
\end{center}
\end{minipage}
\ \ \hspace{0.5cm}
\begin{minipage}[t]{5.4cm}
\begin{center}
\begin{picture}(4.5,4.5)
\put(4.5,0){\scriptsize $x_{\mbox{\tiny I$\!$P}}$}
\put(0,0){\centerline{\epsfysize=4.5cm\epsffile{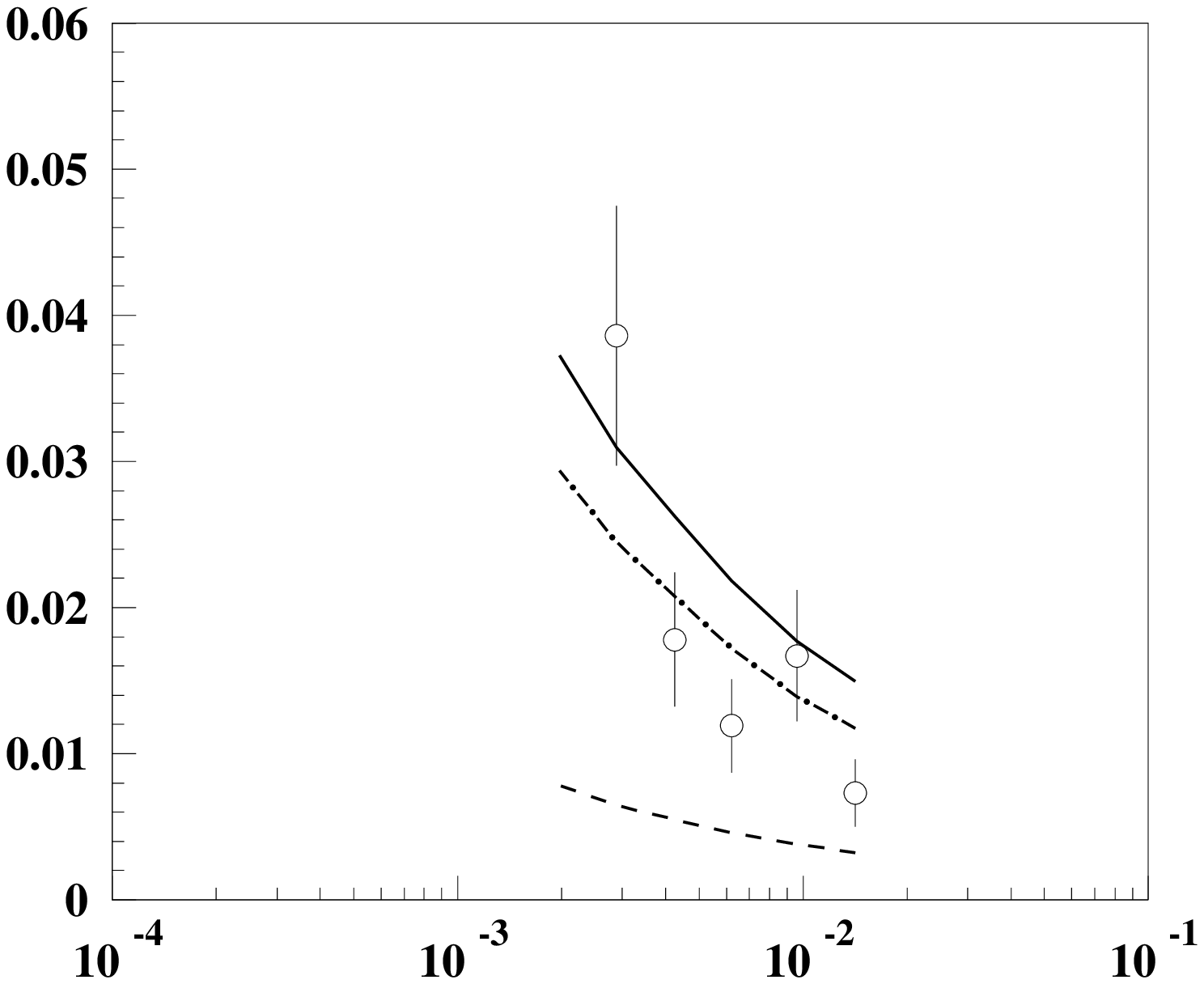}}}
\end{picture}
{\scriptsize \mbox{} \\ \hspace{10mm} (d) $Q^2=60.0$~\mbox{GeV$^2$}, $\beta = 0.938$}
\end{center}
\end{minipage}
\vspace{-0.8cm}
{\caption{\scriptsize \it $x_{\mbox{\tiny I$\!$P}} F_{T}^{D(3)}(x_{\mbox{\tiny I$\!$P}},\beta,Q^2)$ and $x_{\mbox{\tiny I$\!$P}} F_{L}^{D(3)}(x_{\mbox{\tiny I$\!$P}},\beta,Q^2)$ compared with ZEUS~\cite{Zeus1999} results (circles). The transversal structure function (dot-dashed line) fits well the experimental points. Whereas the longitudinal (dashed line) shows a non-vanishing contribution that should be taken into account. Our $x_{\mbox{\tiny I$\!$P}} F_{2}^{D(3)}(x_{\mbox{\tiny I$\!$P}},\beta,Q^2)$ (solid line) calculation overstimates the experimental results. The Monte Carlo integration error is about 10\%.} \label{xpdep}}
\end{center}
\end{figure*}
\end{center}

The $x_{\mbox{\tiny I$\!$P}}$ dependence is plotted in
figure~\ref{xpdep}.  For these plots we took the experimental points
reported by the ZEUS collaboration~\cite{Zeus1999} (circles in
Fig.~\ref{xpdep}).  They have assumed a very small diffractive
longitudinal cross section, indeed $\sigma^{diff}_{\rm L} \ll
\sigma^{diff}_{\rm T}$.  We plotted here the transversal and
longitudinal structure function separately. One can see that although
the longitudinal structure function (dashed line in Fig.~\ref{xpdep})
is con\-siderably smaller than the transversal one (dot-dashed line in
Fig.~\ref{xpdep}), it should not be neglected. From our analysis, one
can observe that although the sum of the longitudinal and transversal
contribution will overestimate the experimental va\-lues of
$x_{\mbox{\tiny I$\!$P}}F_{\rm 2}^{D(3)}$, alone the transversal
structure function fits very good the data and follows the shape 

\newpage

%from f2qbwt.dat

\begin{center}
\begin{figure*}[!hp]
\begin{center}
\unitlength1cm
\begin{minipage}[t]{7.4cm}
\begin{picture}(4.5,4.5)
\put(5.5,0.2){\scriptsize $W$/GeV}
\put(1,2.5){\rotatebox{90}{\scriptsize $d\sigma^{\rm diff}/dt \; / \mbox{nb/GeV$^2$}$}}
\put(0,0){\centerline{\epsfysize=4.5cm\epsffile{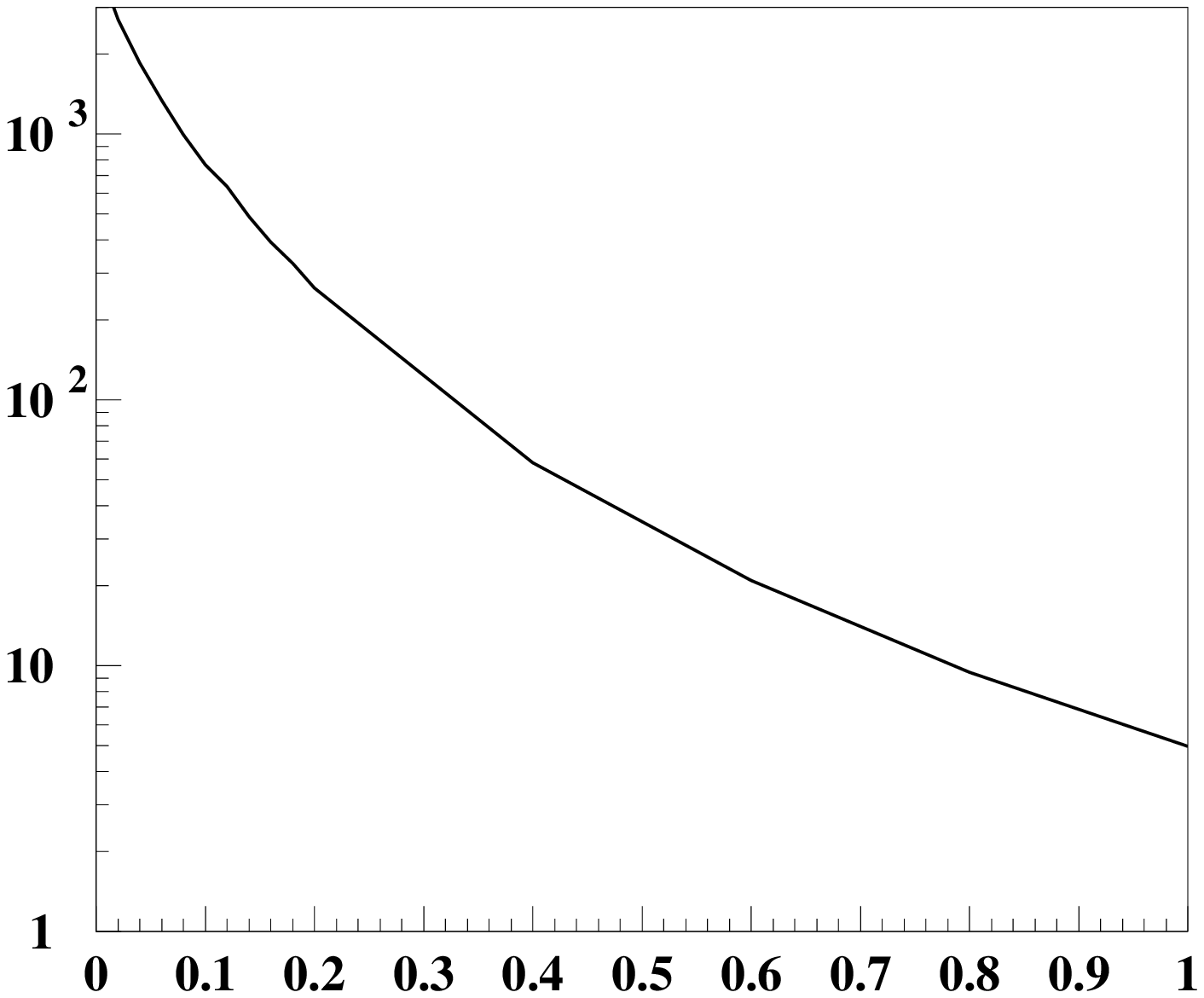}}}
\end{picture}
\vspace{-1cm} {\caption{\scriptsize \it Diffractive differential cross
    section for $W=75.0$ GeV, $Q^2=17.0\mbox{GeV$^2$}$. From this
    curve we obtain the logarithmic slope $B_{\rm D} = 6.55783
    \mbox{GeV$^{-2}$}$ when an exponential behavior in $t$ is
    assumed.}\label{slope}}
\end{minipage}
\ \ \hspace{0.5cm} 
\begin{minipage}[t]{7.4cm}
\begin{picture}(4.5,4.5)
\put(1,4){\rotatebox{90}{\scriptsize ${\sigma^{\rm diff}}/{\sigma^{\rm tot}}$}}
\put(5.5,0.2){\scriptsize $W/$GeV}
\put(0,0){\centerline{\epsfysize=4.5cm\epsffile{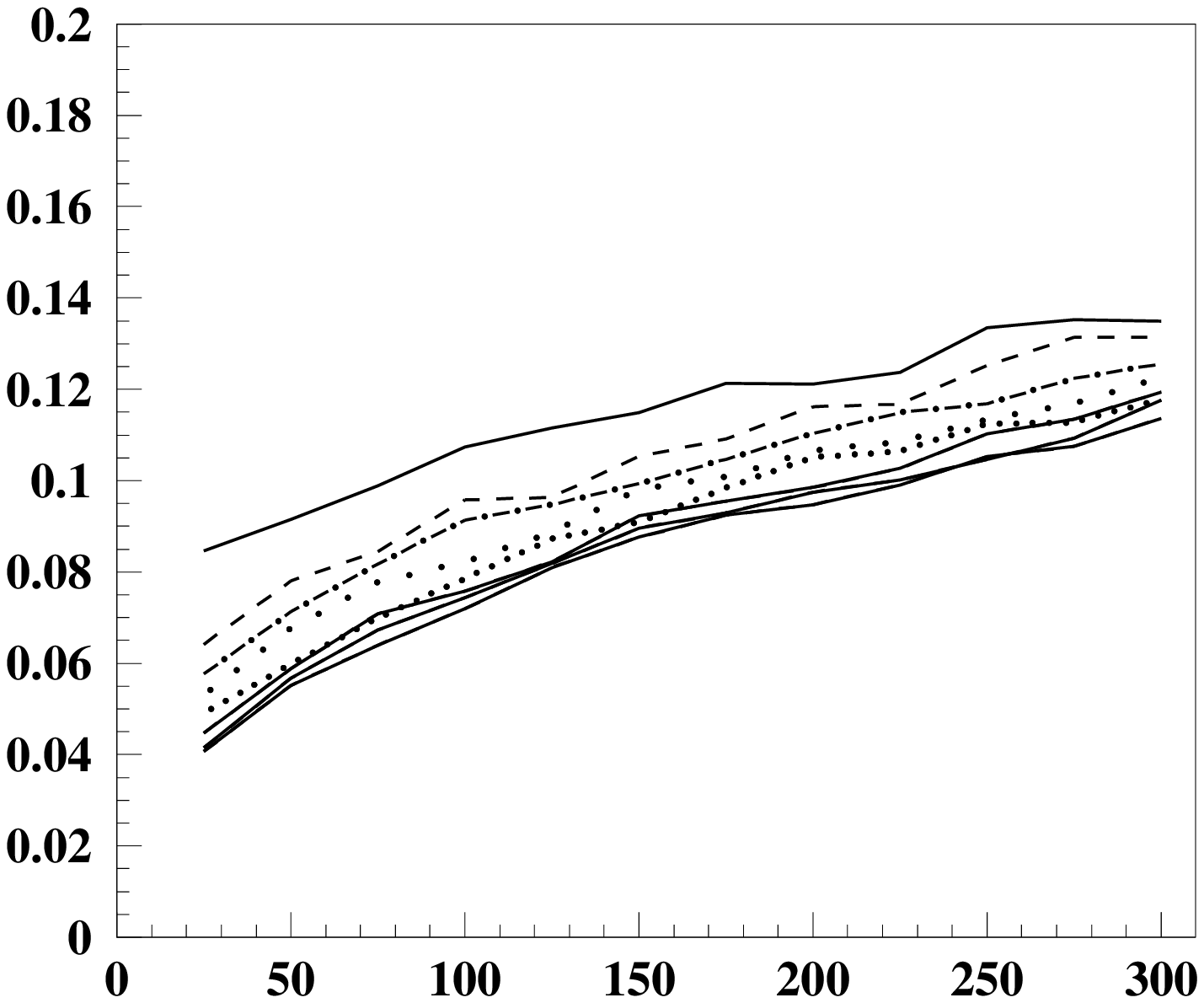}}}
\end{picture}
\vspace{-1cm}
{\caption{\scriptsize \it Ratio of the di\-ffrac\-tive vs. the in\-clu\-sive cross
    section for $Q^2 =$ 5.0 (upper line), 9,0, 13.0, 17.0, 21.0, 25.0,
    29.0, 33.0 (lower line) \mbox{\it GeV$^2$}. In contrast to the
    experiment~\cite{Zeus1999}, which shows a constant value for the
    ratio, our model predicts a growing up ratio with increasing $W$.}
  \label{ratio}}
%\end{center}
\end{minipage}
\end{center}
\end{figure*}
\end{center}

\newpage
\noindent of the curve.  Although our calculation yields directly the $t$
dependent diffractive structure functions $F_{\rm
  L}^{D(4)}(x_{\mbox{\tiny I$\!$P}},\beta,Q^2,t)$ and $F_{\rm
  T}^{D(4)}(x_{\mbox{\tiny I$\!$P}},\beta,Q^2,t)$, a direct comparison
with the experimental va\-lues was not possible. Since the $\beta$
region, where the experimental data are available ($\beta <
0.5$~\cite{ZEUS1998}) involves very small dipoles and perturbative two
gluon exchange should be taken into account.

\section{Diffractive Cross Section in DIS}

The physical picture is exactly the same as in the previous section.
The total cross section is proportional to the imaginary part of the
elastic scattering amplitude of the process $\gamma^{*} + p
\rightarrow \gamma^{*} + p $ according to the optical theorem and
can be written as in~\cite{Nikolaev1991} 

\begin{eqnarray} \label{sigtotalDP}
\lefteqn{\sigma^{\rm tot}_{L,T}(Q^2,W) = \int \frac{d^2r}{4\pi} \int dz |\Psi^{\gamma^{*}}_{L,T}(z,r)|^2} \nonumber \\
&& \times \;\; \sigma(\vec{r},z,W,t=0)
\end{eqnarray}

\noindent where $\Psi^{\gamma^{*}}_{L,T}$ is the wave function for longitudinal and transversal polarized photons~\cite{Dosch1997}. 
The function $\sigma(\vec{r},z,W,t)$ when averaged over the dipole orientation $\varphi$ is the so-called effective dipole-proton cross section

\begin{eqnarray} \label{sigmarho}
\sigma\left(r,z,W,t \right) &=& \int \frac{d\varphi}{2\pi} 2 (2\pi) \int b \, db \, J_{0}\left(\sqrt{|t|}\:b\right) \nonumber \\ 
& & \times \tilde{W}\left(b,\vec{r},z,W,t,\right)
\end{eqnarray}

\noindent where $\tilde{W}\left(b,\vec{r},z,W,t,\right)$ is defined in Eq.~(\ref{functionw}).

On the other hand, writing the phase space Eq.~(\ref{phasespace}) in terms of $t$ and the relative transverse momentum of the quark-antiquark pair $\vec{p}$ 
%\begin{equation} \label{phasespacetp}
%d\Phi_3=\frac{1}{(2\pi)^9}\frac{1}{4W^2}\frac{1}{z(1-z)}dz\;d|t|\;d^2p\;d\varphi,
%\end{equation}
%\noindent 
the differential cross section can be written in terms of $\sigma(\vec{r},z,W,t)$ as 

\begin{eqnarray} \label{diffcross}
\frac{d\sigma^{\rm D}_{L,T}}{dt} &=& \frac{1}{16 \pi}\int \frac{d^2r}{4 \pi} \int dz \left|\Psi^{\gamma^*}_{L,T}\left(r,z\right)\right|^2 \nonumber \\
&&\times \; \sigma^2\left(\vec{r},z,W,t,\right).
\end{eqnarray}

We can relate the differential cross section obtained above Eq.~(\ref{diffcross}) to the diffractive slope $B_{\rm D}$, if we assume an exponential $t$ dependence of the form

\begin{equation} \label{fit}
\frac{d\sigma^{\rm D}}{dt} = A e^{-B_{\rm D} t},
\end{equation}

\noindent where $\sigma^{\rm D}=\sigma^{\rm D}_{L}+\sigma^{\rm D}_{T}$. Doing so, the $t$ integrated diffractive cross section is given by

\begin{equation} \label{tdiffcross}
\sigma^{\rm D}=\left. \frac{1}{B_{\rm D}}\frac{d\sigma^{\rm D}}{dt}\right|_{t=0}.
\end{equation}

Concerning the $t$ dependence, in figure~\ref{slope} we plotted the
diffractive differential cross section Eq.~(\ref{diffcross}) as a
function of $t$. By fitting the solid curve to an exponential function
Eq.~(\ref{fit}), we extracted the logarithmic slope $B_{\rm D}$.  Our
results for the logarithmic slope are comparable with the experimental
data~\cite{ZEUS1998} as shown in the following table.

\vspace{-1cm}
\begin{center}
\begin{table}[!h]
\begin{center}
\begin{tabular}[c]{||l|l||} \hline
\mbox{}& $b$(\mbox{GeV$^{-2})$} \\ \hline
experiment~\cite{ZEUS1998} & $7.2 \pm 1.1$ \\ \hline
MSV (Fig.~\ref{slope})&$6.55783 $ \\ \hline
\end{tabular}
\end{center}
\end{table} 
\end{center}

\vspace{-1cm}

%The absolute value however, should be seen as a prediction of our model, because the statistics of the direct measurements of the $t$ dependence is not good enough.

Figure~\ref{ratio} shows the ratio of the diffractive versus the
inclusive cross section as a function of $W$ for different values of
$Q^2$. Our model predicts a growing up ratio with increasing $W$
independently from $Q^2$. To get this curve, we have integrated
analytically over the whole $\beta$ range.

\end{document}